\documentclass[aps, prd, amsmath, floats, floatfix, twocolumn, nofootinbib, showpacs]{revtex4-1}
\usepackage{graphicx}
\usepackage{color}

\newcommand{\be}{\begin{eqnarray}}
\newcommand{\ee}{\end{eqnarray}}

\begin{document}

\title{Gravitational blueshift from a collapsing object}

\author{Lingyao Kong}

\author{Daniele Malafarina}

\author{Cosimo Bambi}
\email[Corresponding author: ]{bambi@fudan.edu.cn}

\affiliation{Center for Field Theory and Particle Physics \& Department of Physics, Fudan University, 200433 Shanghai, China}

\date{\today}

\begin{abstract}
We discuss a counterintuitive phenomenon of classical general relativity, in 
which a significant fraction of the radiation emitted by a collapsing object and
detected by a distant observer may be blueshifted rather than redshifted. 
The key-point is that when the radiation propagates inside the collapsing body 
it is blueshifted, and this time interval may be sufficiently long for the effect to 
be larger than the later redshift due to the propagation in the vacuum exterior, 
from the surface of the body to the distant observer. Unfortunately, the phenomenon 
can unlikely have direct observational implications, but it is interesting by itself 
as a pure relativistic effect.
\end{abstract}

\pacs{04.20.-q, 97.60.-s, 04.40.Nr}

\maketitle


The phenomenon of photon redshift is well known in cosmology. A photon 
propagating in an expanding Universe is redshifted by the factor $a(t_1)/a(t_2)$ 
from the time $t_1$ to the time $t_2$, where $a(t)$ is the scale factor of the 
Friedmann-Robertson-Walker (FRW) metric. If the Universe were contracting, the 
photon would be blueshifted. In this note, we discuss the shift experienced 
by the photons emitted by a collapsing body. In the homogeneous case,
the interior metric in comoving coordinates can be written as the time reversal of the
FRW metric, therefore it is the same as the one of a contracting Universe and photons are thus blueshifted. The crucial point is whether such photons can retain some blueshift when they reach far-away observers.
For a spherically symmetric system, the exterior vacuum spacetime is 
described by the Schwarzschild metric, and it is well known that every photon 
will be redshifted when it propagates from the surface of the object to the distant observer. 
The counterintuitive result we show here is that a significant fraction of the photons emitted by 
the collapsing object may eventually be seen as blueshifted even by the distant
observer; that is, the blueshift occurring inside the body may exceed the
subsequent redshift in the vacuum.

For the sake of simplicity, we study the radiation emitted by the collapse
of a spherically symmetric cloud of dust. We consider both the homogeneous
(constant energy density) and inhomogeneous cases, the well known Lema\`itre-Tolman-Bondi (LTB) model
\cite{LTB}. 
As it follows from Birkhoff's theorem, the exterior metric is described by the Schwarzschild
solution and the line element is
\be
\hspace{-0.4cm}
ds^2_{\rm out} = - \left(1 - \frac{2M_{\rm Sch}}{R}\right) dT^2 
+ \frac{dR^2}{\left(1 - \frac{2M_{\rm Sch}}{R}\right)}
+ R^2 d\Omega^2 \, .
\ee
Here $d\Omega^2 = d\theta^2 + \sin^2\theta d\phi^2$ represents the line 
element on the unit two-sphere and $R > R_{\rm b}(T)$, where $R_{\rm b}(T)$ is
the radial coordinate of the boundary radius of the collapsing body. Since the spacetime 
is spherically symmetric, without loss of generality we can restrict our study 
to the equatorial plane $\theta = \pi/2$. The exterior spacetime is also static and
axisymmetric, and we thus have two constants of motion; that is, the
energy $E$ and the angular momentum $L$
\be
E = \left(1 - \frac{2M_{\rm Sch}}{R}\right) \frac{dT}{d\lambda} \, , 
\quad L = R^2 \frac{d\phi}{d\lambda} \, ,
\ee
where $\lambda$ is an affine parameter. We define the impact parameter as $b = L/E$ and it 
corresponds to the radial coordinate on the image plane of the distant observer. 
For null geodesics, $g_{\mu\nu} (dx^\nu/d\lambda)(dx^\nu/d\lambda) = 0$ and 
the motion of the photon in the vacuum spacetime with respect to the proper time 
of the distant observer is described by
\be\label{eq-out}
\left(\frac{dR}{dT}\right)^2 = \left(1 - \frac{2M_{\rm Sch}}{R}\right)^2
- \left(1 - \frac{2M_{\rm Sch}}{R}\right)^3 \frac{b^2}{R^2} \, .
\ee

The spacetime inside the collapsing body is described by a dynamical interior solution
(see e.g. Ref.~\cite{bmm}). 
The most general spherically symmetric metric describing a collapsing 
cloud of matter in comoving coordinates is given by
\begin{equation}\label{eq1}
ds^2_{\rm in} = -e^{2\nu}dt^2+\frac{\rho'^2}{G}dr^2
+\rho^2d\Omega^2 \; ,
\end{equation}
where $\nu$, $\rho$, and $G$ are functions of $t$ and $r$. The energy-momentum 
tensor is given by 
\begin{equation}
T^\mu_\nu={\rm diag}\{\varepsilon(r,t), p_r(r,t), p_\theta(r,t), p_\theta(r,t) \} \; ,
\end{equation} 
and Einstein's equations relate the metric functions to the matter content
\begin{eqnarray}\label{rho}
  p_r &=&-\frac{\dot{F}}{\rho^2\dot{\rho}} \; , \: 
  \varepsilon=\frac{F'}{\rho^2\rho'} \; , \\ \label{nu}
  \nu'&=&2\,\frac{p_\theta-p_r}{\varepsilon+p_r}\frac{\rho'}{\rho}
  -\frac{p_r'}{\varepsilon+p_r} \; ,\\ \label{Gdot}
  \dot{G}&=&2\,\frac{\nu'}{\rho'}\dot{\rho}\,G \; ,
\end{eqnarray}
where the $'$ denotes a derivative with respect to $r$, and the $\dot{}$ denotes a 
derivative with respect to $t$. The function $F(r,t)$ is called Misner-Sharp mass, 
and is
\begin{equation}\label{misner}
F=\rho(1-G+e^{-2\nu}\dot{\rho}^2) \; .
\end{equation}

The case of marginally bound collapse of a cloud of dust is particularly simple.
Since $p_r = p_\theta = 0$, from the first of Eq.~\eqref{rho} it follows that $F$ is a function of $r$ only and the matching to the exterior vacuum Schwarzschild spacetime is always possible
\cite{matching}. 
Furthermore, Eq.~\eqref{nu} reduces to $\nu'=0$ and we can always choose the time coordinate in such a way that
$\nu = 0$. Integration of Eq.~\eqref{Gdot}  is then trivial and gives $G = 1+f(r)$ and in the marginally bound collapse case we shall take the free integration function $f$ to be zero. 

The whole system has a gauge degree of freedom that 
can be fixed by setting the scale at a certain time. The usual prescription is that 
the area radius $\rho(r,t)$ is set equal to the comoving radius $r$ at the initial 
time $t_{\rm i}=0$, $\rho(r,0)=r$. We can then introduce a scale function $a$
\begin{equation}
\rho(r,t)=ra(r,t) \; ,
\end{equation}
that will go from 1, at the initial time, to 0, at the time of the formation of the 
singularity. The condition to describe collapse is thus given by $\dot{a}<0$. 
The regularity of the energy density at the initial time, as seen from Eq.~\eqref{rho}, 
requires that $F(r)=r^3M(r)$, with $M(r) = \sum_{n = 0}^{\infty} M_n r^n$. 
From the equation for $\varepsilon$ in~(\ref{rho}), it follows that the Misner-Sharp 
mass at the boundary $r_{\rm b}$ is two times the gravitational mass of the body
and therefore
\be\label{eq-ms}
2 M_{\rm Sch} = F(r_{\rm b}) = r_{\rm b}^3 M(r_{\rm b}) \, .
\ee
The energy density can then be written from the second of Eq.~\eqref{rho} as 
\be
\varepsilon = \frac{3M + rM'}{a^2 (a + ra')}\, .
\ee
The Misner-Sharp mass, Eq.~\eqref{misner}, takes the form of an equation 
of motion
\begin{equation}\label{motion}
\dot{a}=-\sqrt{\frac{M}{a}} \; ,
\end{equation}
with the minus sign chosen in order to describe a collapse. The integration of 
Eq.~\eqref{motion} is straightforward and gives
\begin{equation}\label{a1}
a(r,t)=\left(1-\frac{3}{2}\sqrt{M}t\right)^{2/3} \; .
\end{equation}

On the surface of the collapsing object we have $\rho_{\rm b}(t) =\rho(t,r_{\rm b}) = R_{\rm b}(t)$
and $ds^2_{\rm in} = ds^2_{\rm out}$, so
\be\label{eq1}
dt^2 = \left(1 - \frac{2M_{\rm Sch}}{R_{\rm b}}\right) dT^2 - 
\frac{dR_{\rm b}^2}{\left(1 - \frac{2M_{\rm Sch}}{R_{\rm b}}\right)} \, .
\ee
Marginally bound collapse describes infalling particles that have zero initial velocity at infinity. The surface of the body thus follows a marginally 
bound time-like geodesic in the Schwarzschild spacetime
\be\label{eq2}
\frac{d\rho_{\rm b}}{dt} = 
\frac{dR_{\rm b}}{dt} =
- \sqrt{\frac{2M_{\rm Sch}}{R_{\rm b}}} \, ,
\ee
where $t$ is the proper time of the collapsing body and coincides with
the comoving time coordinate of the interior solution. From Eqs.~(\ref{eq1}) and (\ref{eq2}), 
we find the relation between the time coordinates $t$ and $T$:
\be
\left(\frac{dT}{dt}\right)^2 &=& \left(1 - 
\frac{2M_{\rm Sch}}{R_{\rm b}}\right)^{-2}
\left[1 - \frac{2M_{\rm Sch}}{R_{\rm b}} + 
\left(\frac{dR_{\rm b}}{dt}\right)^2\right] = \nonumber\\
&=& \left(1 - \frac{2M_{\rm Sch}}{R_{\rm b}}\right)^{-2} \, .
\ee
The motion of the surface of the body with respect to the proper time of the
distant observer is given by
\be
\frac{dR_{\rm b}}{dT} = \frac{d\rho_{\rm b}}{dt} \frac{dt}{dT} =
- \sqrt{\frac{2M_{\rm Sch}}{R_{\rm b}}} 
\left(1 - \frac{2M_{\rm Sch}}{R_{\rm b}}\right) \, ,
\ee
which can be integrated to get the time $T$ at which the radial coordinate
of the collapsing object has radius $R_{\rm b}$
\be\label{eq-t}
T &=& T_0 - \frac{2}{3} \sqrt{\frac{R_{\rm b}}{2M_{\rm Sch}}} 
\left(R_{\rm b} + 6M_{\rm Sch}\right) 
+ \nonumber\\ && \hspace{0.2cm}
+ 2M_{\rm Sch} \ln \left(\frac{\sqrt{R_{\rm b}} 
+ \sqrt{2M_{\rm Sch}}}{\sqrt{R_{\rm b}} - \sqrt{2M_{\rm Sch}}}\right) \, .
\ee

To compute the spectrum of the collapsing body seen by the distant observer,
we can proceed as follows. We consider the photons on the image plane of the 
distant observer, which are characterized by their impact parameter $b$, and
we integrate backward in time the trajectory of the photon from the distant 
observer to the surface of the body with the help of Eqs.~(\ref{eq-out}) and (\ref{eq-t}).
After the photon hits the surface, we follow its propagation backward in time in the interior 
solution. Inside the collapsing body, the metric is axisymmetric but not stationary, 
and therefore the photon angular momentum $L$ is conserved but the photon 
energy $E$ is not. We use again the fact that we are considering light-like trajectories,
so $ds^2 = 0$, and we write the equations for the coordinates $t$ and $r$
\be
\frac{d^{2}r}{d\lambda^{2}} &=&
 - \frac{2\dot{\rho'}}{\rho'}\frac{dr}{d\lambda}
\sqrt{\rho'^2\left(\frac{dr}{d\lambda}\right)^2 + \frac{L^2}{\rho^2}} 
+ \nonumber\\ &&-
\frac{\rho''}{\rho'}\left(\frac{dr}{d\lambda}\right)^2 
+ \frac{1}{\rho'}\frac{L^{2}}{\rho^{3}}\, , \\
\frac{dt}{d\lambda} &=&
\sqrt{\rho'^2\left(\frac{dr}{d\lambda}\right)^2+
\frac{L^{2}}{\rho^{2}}}\, .
\ee

The spectrum at the time $T$ measured by the distant observer is given by
\be
I(T,\nu_{\rm obs}) = \int 2\pi b db \, \int_\gamma g^3 j dl \, ,
\ee
where $\gamma$ is the photon's path, $j$ is the emissivity per unit volume in the rest 
frame of the emitter, $g$ is the gravitational redshift
\be
g = \frac{\nu_{\rm obs}}{\nu_{\rm e}} = 
\frac{k_\mu v^{\mu}_{\rm obs}}{k_\nu v^{\nu}_{\rm e}} = 
\frac{E}{\frac{dt}{d\lambda}} \, ,
\ee
$\nu_{\rm obs}$ is the photon frequency as measured by the distant observer,
$\nu_{\rm e}$ is the photon frequency with respect to the emitter, 
$v^{\mu}_{\rm obs}=(1,0,0,0)$ is the 4-velocity of the distant observer, 
$v^{\mu}_{\rm e}=(1,0,0,0)$ is the 4-velocity of the emitter, and $k^\mu$ is the
4-momentum of the photon. For simplicity, here we assume that the emission is 
monochromatic with rest-frame frequency $\nu_\star$ and proportional to the 
square of the energy density $\varepsilon$ (as we may expect in a two-body 
collision)
\be
j = \left\{
\begin{array}{ll}
0 & \text{if } R > R_{\rm b} \, , \\
\varepsilon^2 \delta 
\left(\nu_{\rm e} - \nu_\star \right) 
& \text{if } R < R_{\rm b} \, .
\end{array} \right.
\ee
Note that $dl$ is the proper length in the rest-frame of the emitter and in our model it turns out to be equal to $dt$ 
\be
dl = \sqrt{^3g_{ij} \frac{dx^i}{d\lambda}\frac{dx^j}{d\lambda}}
d\lambda = dt \, .
\ee

If we use
\be
\frac{dR_{\rm b}}{d\lambda} = \dot{R}_{\rm b}\frac{dt}{d\lambda} 
+ R'_{\rm b} \frac{dr}{d\lambda} 
\ee
and
\be
(ds^2_{\rm in})_{r_{\rm b}} = 0 &\Rightarrow&
\frac{L^2}{R_{\rm b}^2} = \left(\frac{dt}{d\lambda}\right)^2
- R'^2_{\rm b} \left(\frac{dr}{d\lambda}\right)^2
\ee
in the equation $(ds^2_{\rm out})_{r_{\rm b}} = 0$, that is
\begin{widetext}
\be
\label{dsout}
\left(\frac{dT}{d\lambda}\right)^2 =
\left(1 - \frac{2M_{\rm Sch}}{R_{\rm b}}\right)^{-2}
\left[\left(\frac{dR_{\rm b}}{d\lambda}\right)^2 + 
\left(1 - \frac{2M_{\rm Sch}}{R_{\rm b}}\right)
\frac{L^2}{R^2_{\rm b}}\right] \, ,
\ee
we find (recalling that $\dot{R}_{\rm b} = - \sqrt{2 M_{\rm Sch}/R_{\rm b}}$)
\be
\left(\frac{dT}{d\lambda}\right)^2 &=&
\left(1 - \frac{2M_{\rm Sch}}{R_{\rm b}}\right)^{-2}
\left\{\left(\dot{R}_{\rm b}\frac{dt}{d\lambda} 
+ R'_{\rm b} \frac{dr}{d\lambda}\right)^2 + 
\left(1 - \frac{2M_{\rm Sch}}{R_{\rm b}}\right)
\left[  \left(\frac{dt}{d\lambda}\right)^2
- R'^2_{\rm b} \left(\frac{dr}{d\lambda}\right)^2 \right]\right\} = \nonumber\\
&=&\left(1 - \frac{2M_{\rm Sch}}{R_{\rm b}}\right)^{-2}
\left(\frac{dt}{d\lambda} - \sqrt{\frac{2M_{\rm Sch}}{R_{\rm b}}} R'_{\rm b} 
\frac{dr}{d\lambda} \right)^2 \, \\
\frac{dT}{d\lambda} &=& \left(1 - \frac{2M_{\rm Sch}}{R_{\rm b}}\right)^{-1}
\left[\frac{dt}{d\lambda} - \sqrt{\frac{2M_{\rm Sch}}{R_{\rm b}}}  
\left( \frac{dR_{\rm b}}{d\lambda} - \dot{R}_{\rm b}\frac{dt}{d\lambda} \right) \right] = 
\frac{dt}{d\lambda}
- \left(1 - \frac{2M_{\rm Sch}}{R_{\rm b}}\right)^{-1}
\sqrt{\frac{2M_{\rm Sch}}{R_{\rm b}}} \frac{dR_{\rm b}}{d\lambda} \, .
\label{dtdlrb}
\ee
From Eq.~(\ref{dsout}) we have
\be
\left(\frac{dR_{\rm b}}{d\lambda}\right)^2 =
\left(1 - \frac{2M_{\rm Sch}}{R_{\rm b}}\right)^2
\left(\frac{dT}{d\lambda}\right)^2
- \left(1 - \frac{2M_{\rm Sch}}{R_{\rm b}}\right)
\frac{L^2}{R^2_{\rm b}} =
E^2 \left[ 1 - \left(1 - \frac{2M_{\rm Sch}}{R_{\rm b}}\right)
\frac{b^2}{R^2_{\rm b}} \right] \, .
\ee
If we plug this expression for $dR_{\rm b}/d\lambda$ into Eq.~(\ref{dtdlrb}), 
we get $dt/d\lambda$ at the boundary
\be
\left(\frac{dt}{d\lambda}\right)_{r_{\rm b}} 
&=& \frac{E}{\left(1 - \frac{2M_{\rm Sch}}{R_{\rm b}}\right)} 
 \left(1 + \sqrt{\frac{2M_{\rm Sch}}{R_{\rm b}}} \sqrt{1 - \left(1 - 
\frac{2M_{\rm Sch}}{R_{\rm b}}\right)\frac{b^2}{R_{\rm b}^2}}\right) \, .
\label{eq-tra}
\ee
\end{widetext}
At infinity, the photon energy is $E$.
In Eq.~(\ref{eq-tra}), we have the usual gravitational redshift of the Schwarzschild metric,
the term $\left(1 - \frac{2M_{\rm Sch}}{R_{\rm b}}\right)^{-1}$, plus a correction due 
to the non-vanishing velocity of the surface of the collapsing body (Doppler redshift).
Eq.~(\ref{eq-tra}) relates the photon energy between the interior and the exterior
solutions, and it is therefore a crucial ingredient in the final result. Let us note that
such an expression was derived in Ref.~\cite{nakao} following a different approach.

\begin{figure*}
\begin{center}
\hspace{-0.5cm}
\includegraphics[type=pdf,ext=.pdf,read=.pdf,width=6cm]{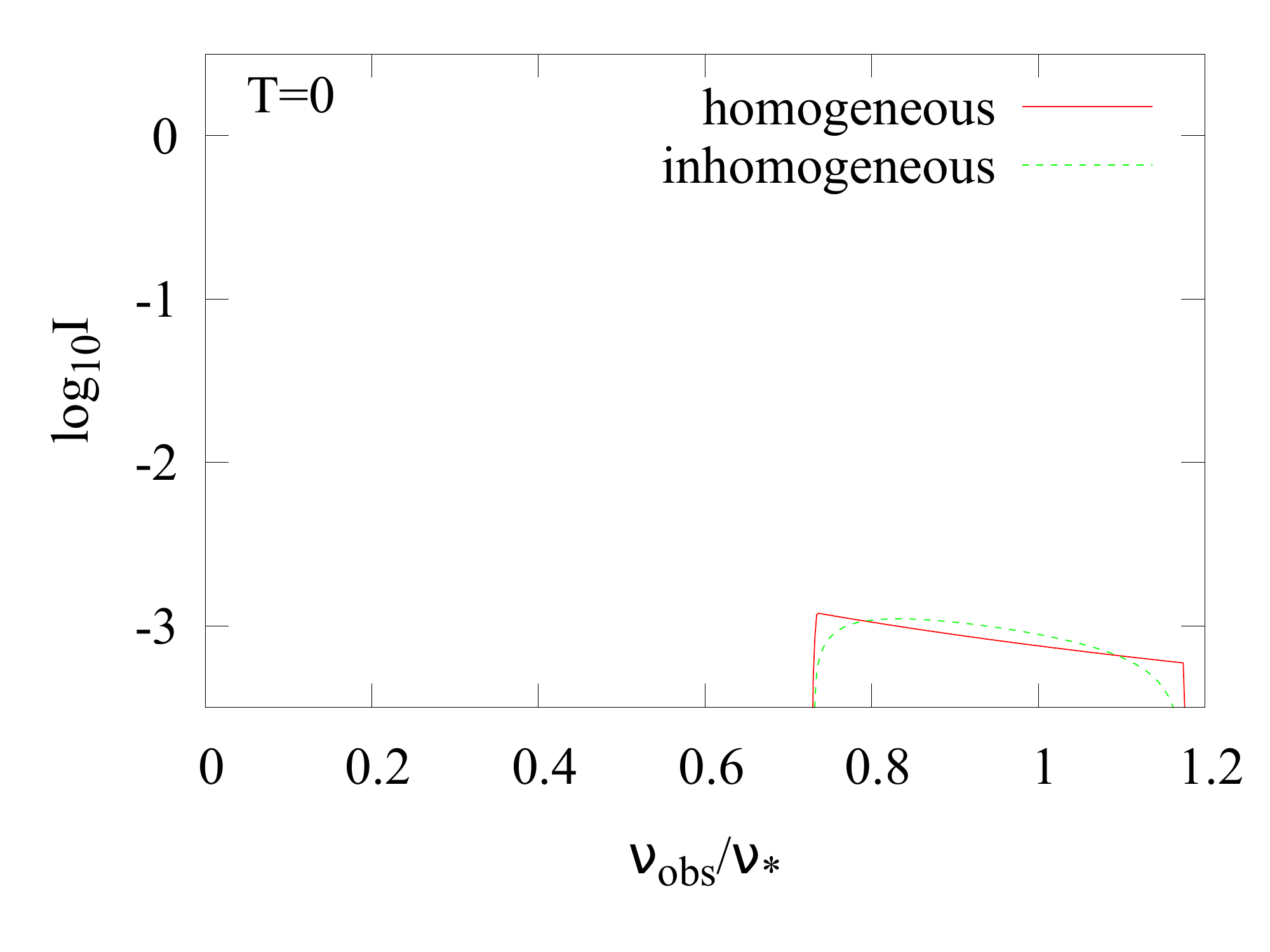}
\includegraphics[type=pdf,ext=.pdf,read=.pdf,width=6cm]{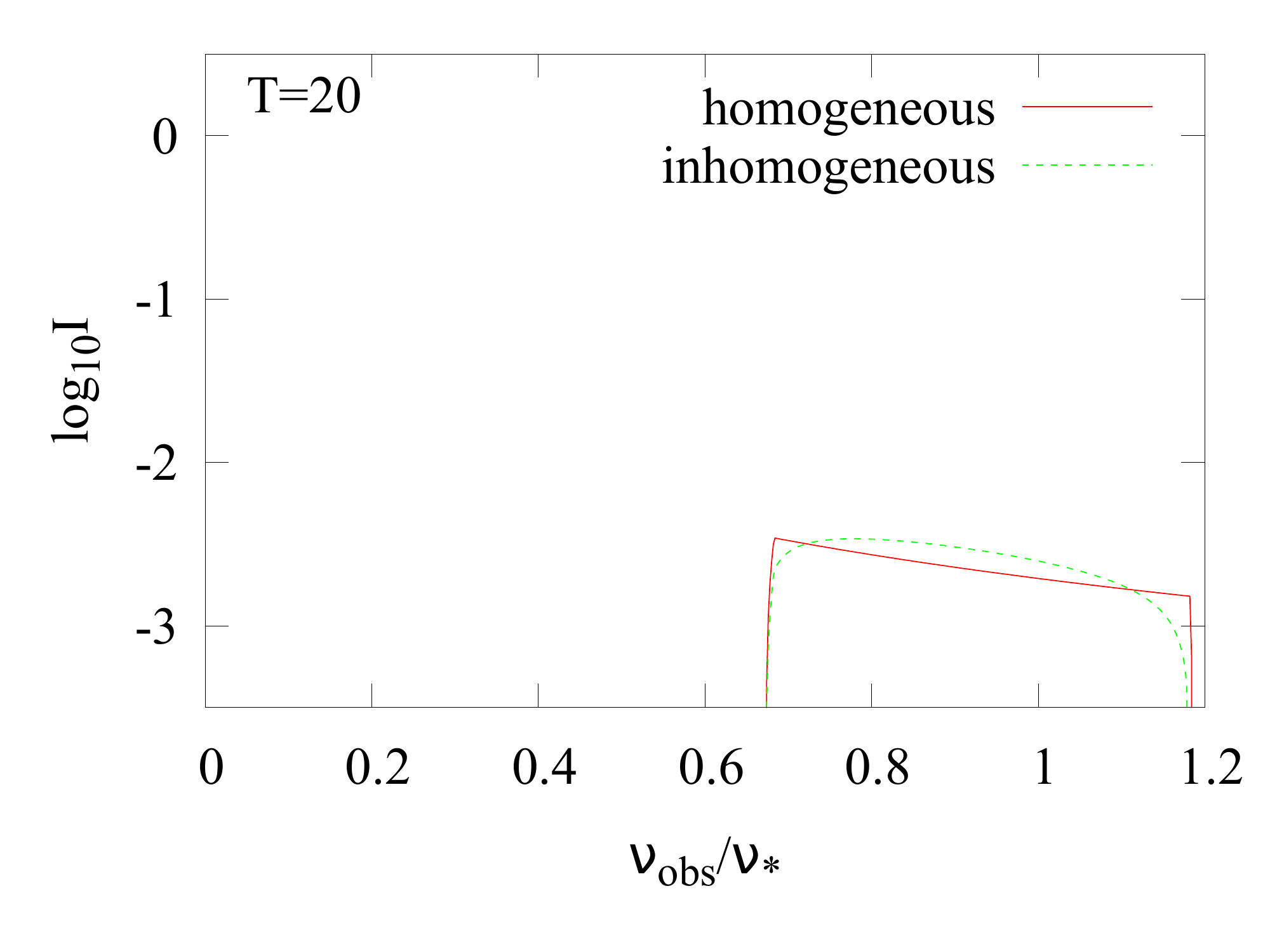}
\includegraphics[type=pdf,ext=.pdf,read=.pdf,width=6cm]{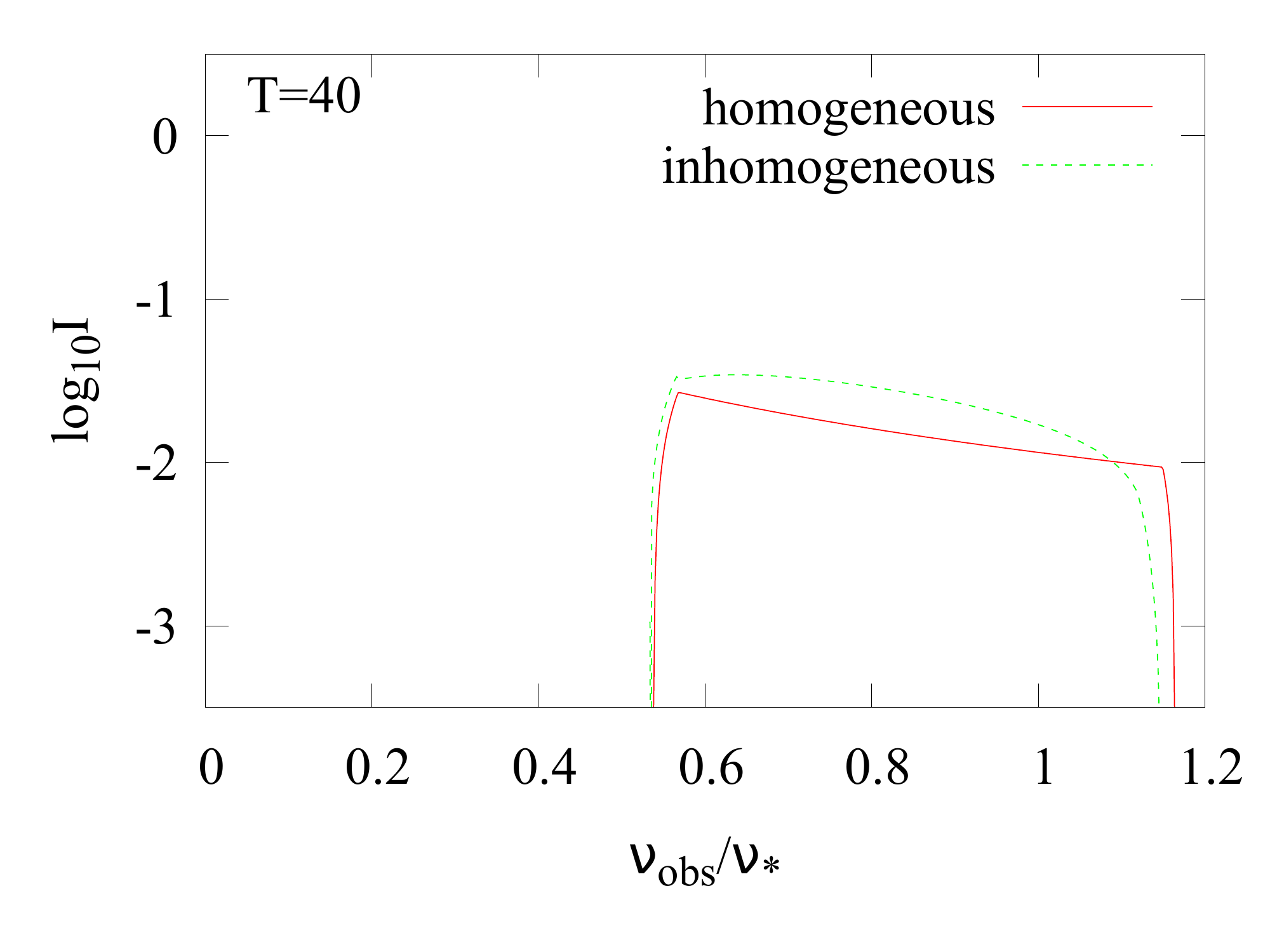} \\
\vspace{0.5cm}
\hspace{-0.5cm}
\includegraphics[type=pdf,ext=.pdf,read=.pdf,width=6cm]{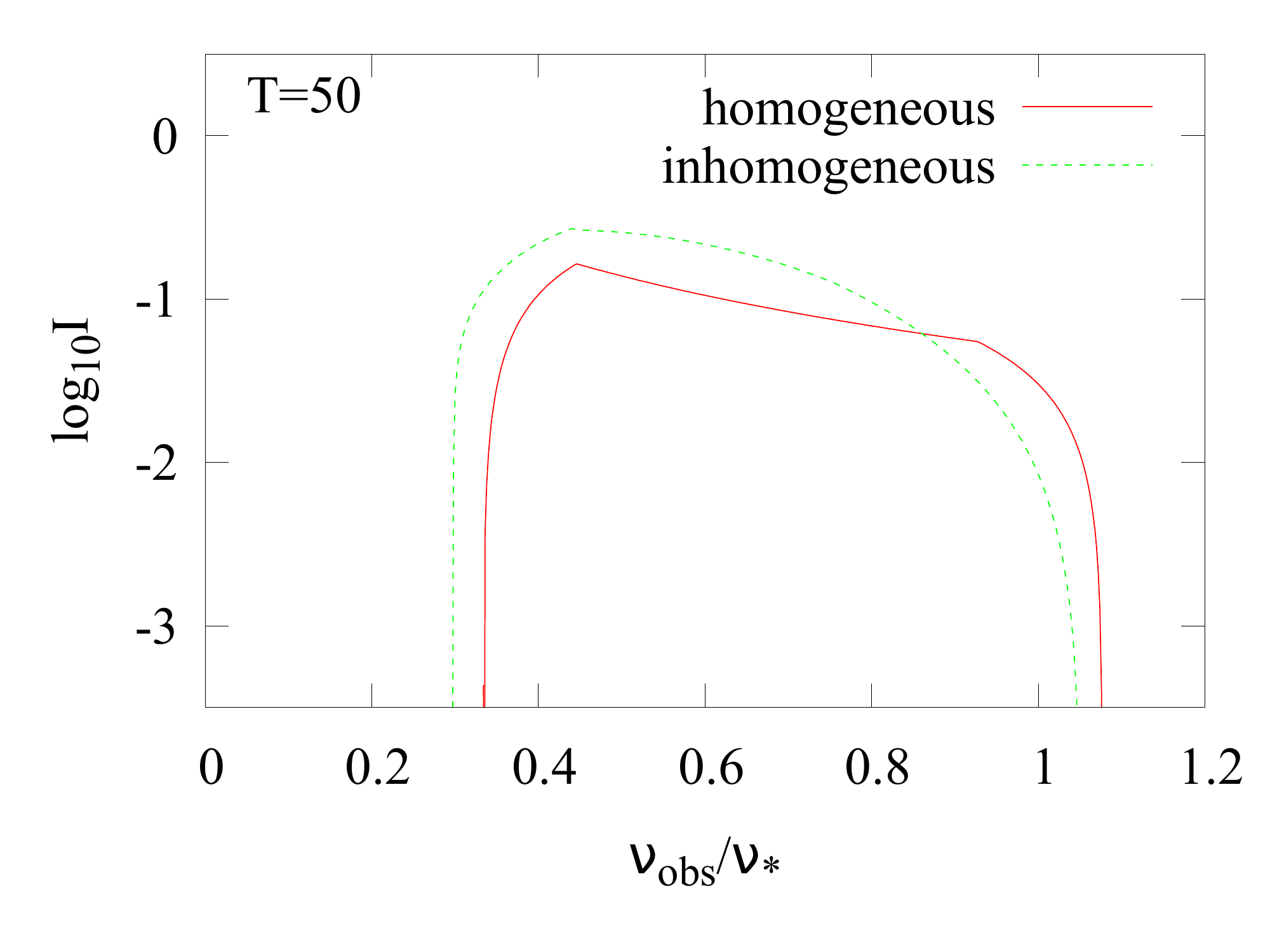}
\includegraphics[type=pdf,ext=.pdf,read=.pdf,width=6cm]{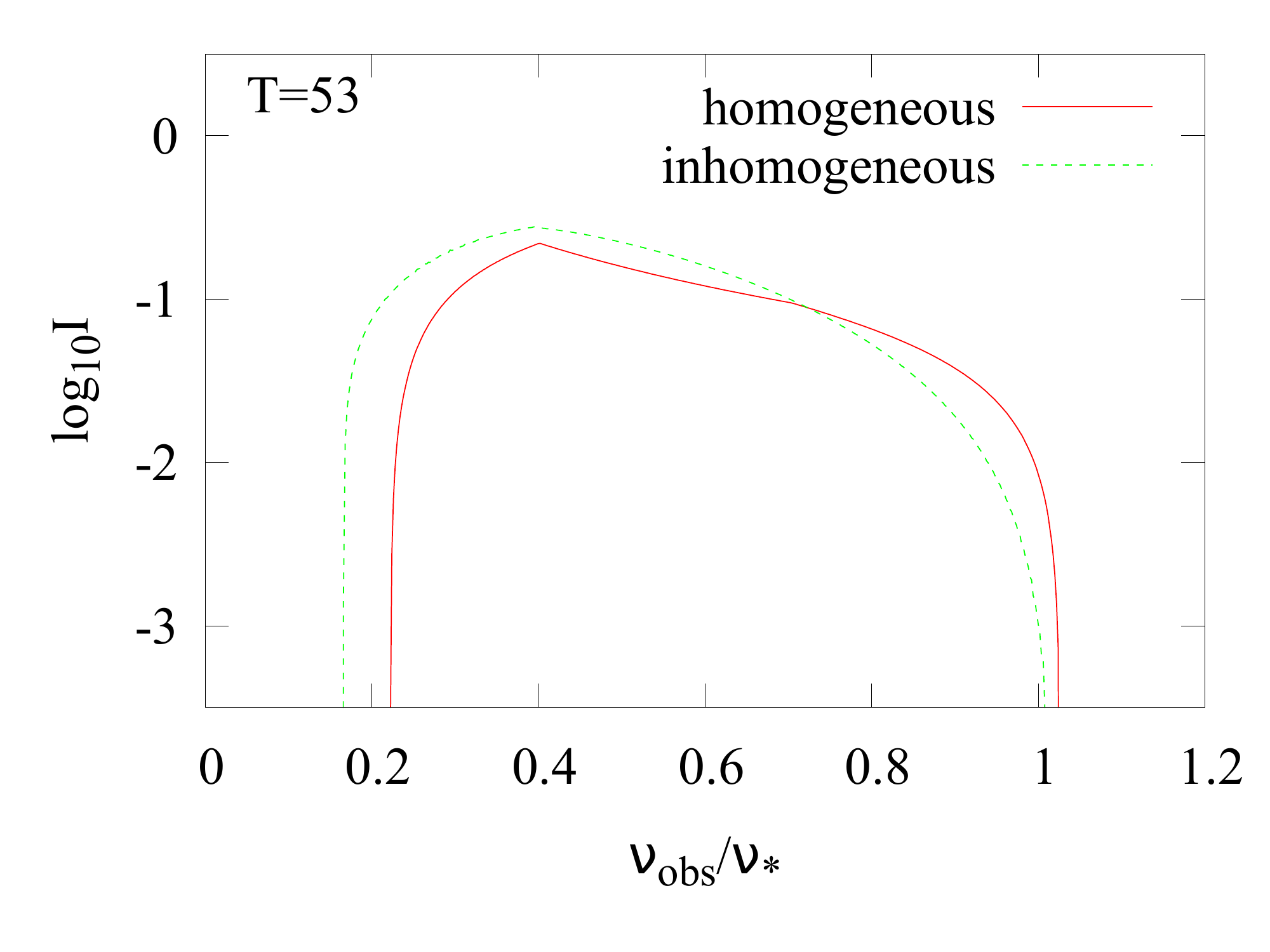}
\includegraphics[type=pdf,ext=.pdf,read=.pdf,width=6cm]{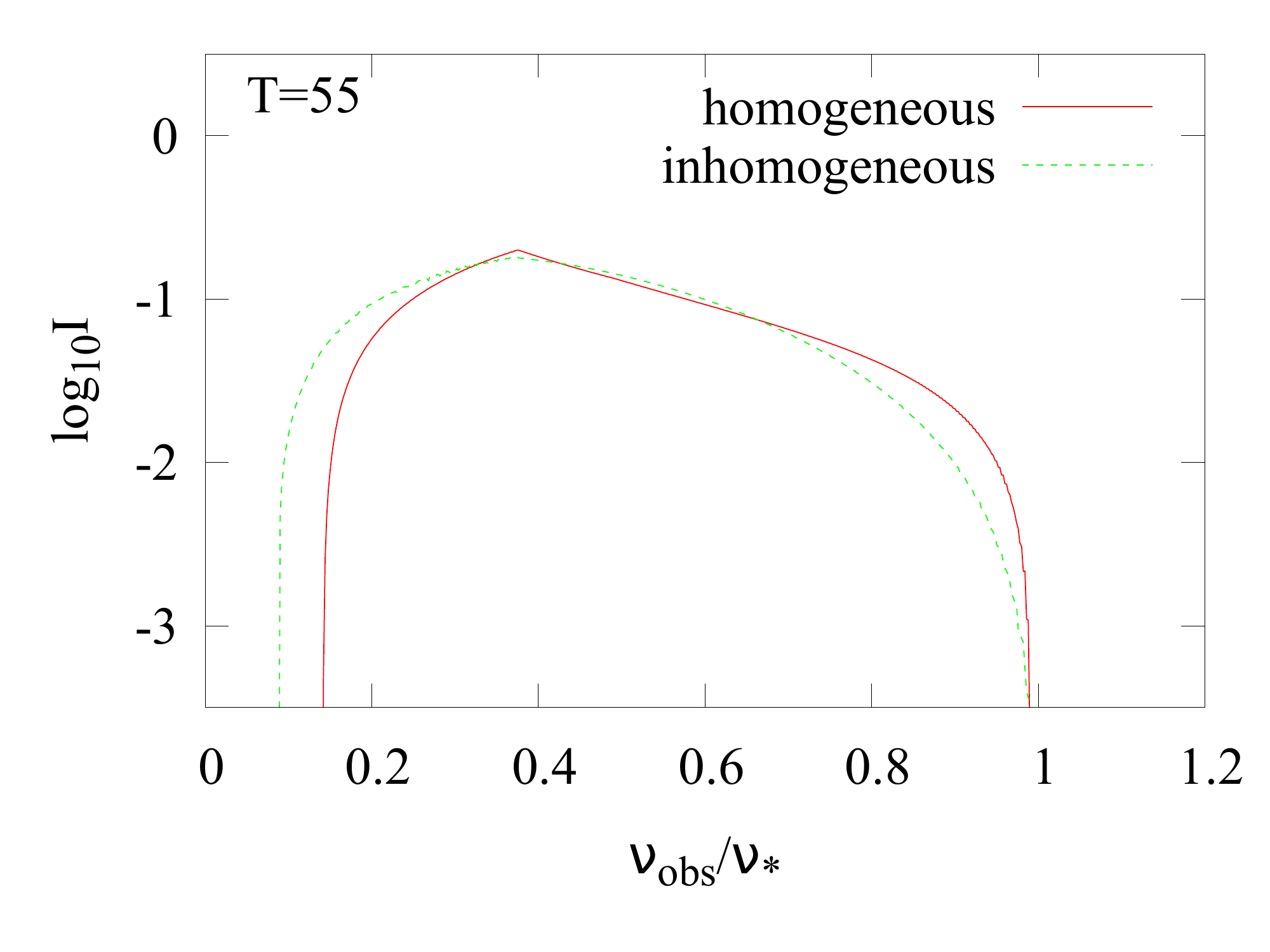}
\end{center}
\caption{Spectra of a homogeneous (red solid line) and inhomogeneous 
(green dashed line) collapsing dust clouds at the time $T = 0$ (top left panel), 
$T = 20$ (top central panel), $T = 40$ (top right panel), $T = 50$ (bottom left 
panel), $T = 53$ (bottom central panel), and $T = 55$ (bottom right panel). Time 
$T$ in units $2M_{\rm Sch} = 1$; intensity $I$ in arbitrary units.}
\label{f2}
\end{figure*}

\begin{figure*}
\begin{center}
\hspace{-0.5cm}
\includegraphics[type=pdf,ext=.pdf,read=.pdf,width=6.8cm]{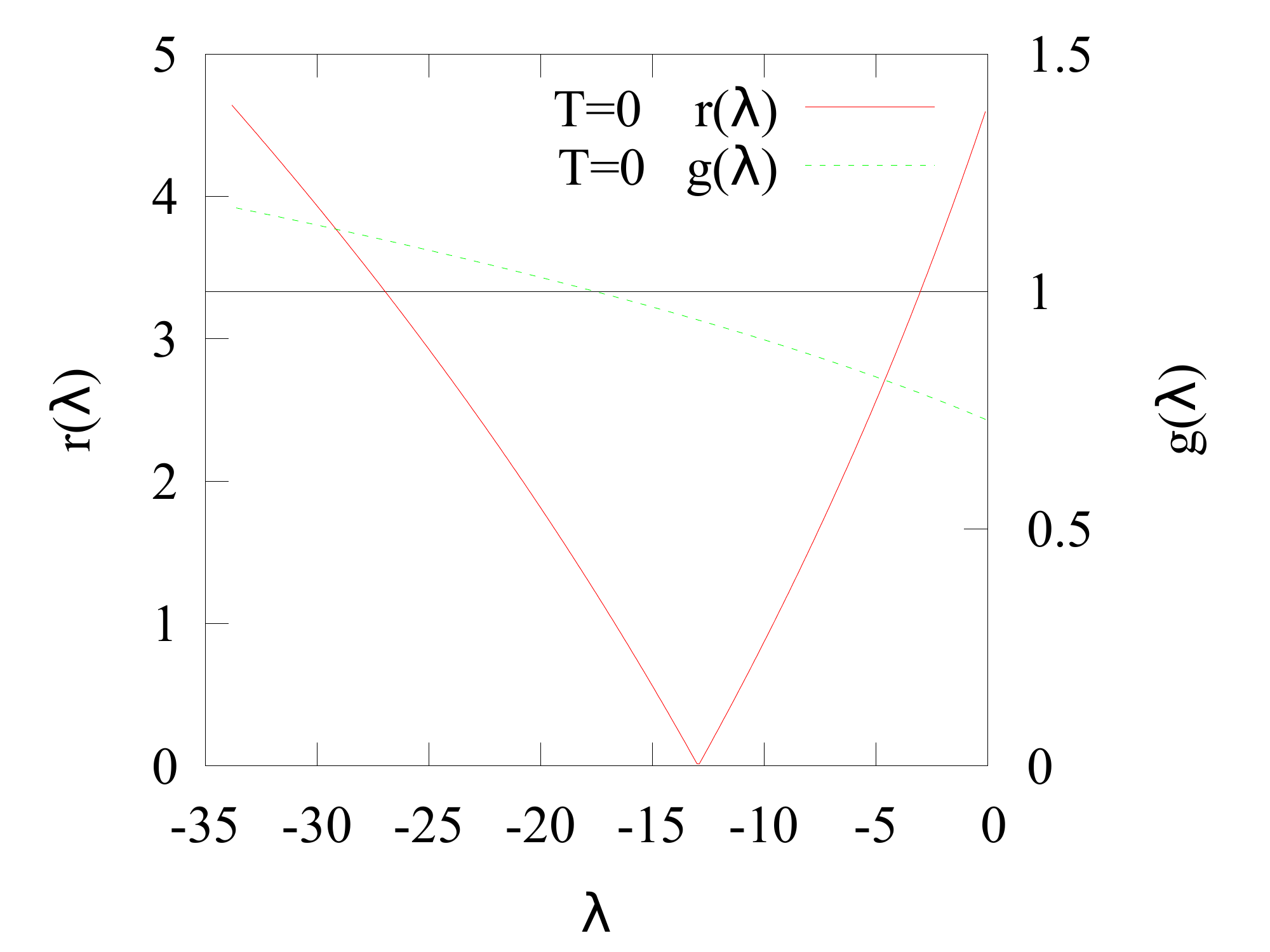}
\hspace{0.8cm}
\includegraphics[type=pdf,ext=.pdf,read=.pdf,width=6.8cm]{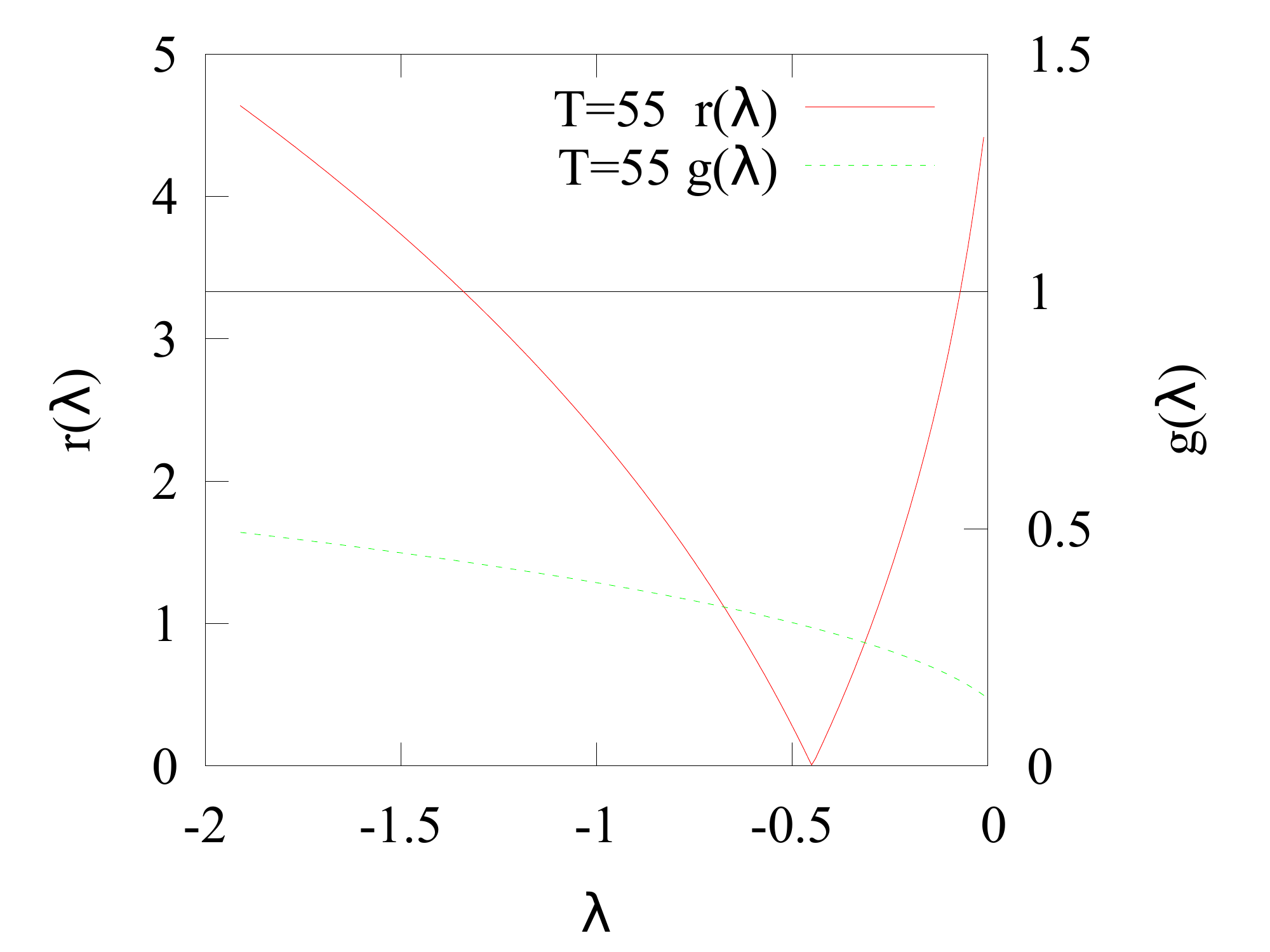}
\end{center}
\caption{Comoving radial coordinate of the point of the emission of the photon
inside the collapsing object (red solid line) and the photon redshift factor $g$ 
(green dashed line) as a function of the affine parameter $\lambda$ for photons
with impact parameter $b=0$ and reaching the distant observer at the time
$T = 0$ (left panel) and $T = 55$ (right panel). Photons emitted at $\lambda=0$
have $r = r_{\rm b}$ and they are affected only by the gravitational redshift in
the Schwarzschild spacetime and by the Doppler redshift due to the non-vanishing
velocity of the surface of the collapsing body. Photons emitted at $r = r_{\rm b}$ 
and finite $\lambda$ ($\lambda \approx 35$ for $T=0$ and 2 for $T=55$) cross
the whole body and exit from the opposite side: they experience the maximum 
gravitational blueshift and arrive at the distant observer with the highest energies.
Time $T$ in units $2M_{\rm Sch} = 1$. Homogeneous collapse model.}
\label{f3}
\end{figure*}

Fig.~\ref{f2} shows the spectra of the collapsing body for different times in the 
case of a homogeneous and inhomogeneous cloud of dust. In the homogeneous
case, we have $M(r) = M_0$, where $M_0$ is a constant determined by the
condition $r_{\rm b}^3 M_0 = 2 M_{\rm Sch}$. In the inhomogeneous case,
we have
\be
M(r) = M_0 + M_2 r^2 \, ,
\ee
and we have chosen $M_0 = 0.01$ and $M_2 = -0.00015$ ($M_2 < 0$ to have a higher density
at the center of the body and $M_0 + M_2 r^2_{\rm b} > 0$ to have positive density). Especially at the beginning of the collapse, when the boundary is larger, we have
a quite unexpected fraction of radiation with frequency $\nu_{\rm obs} > \nu_\star$;
that is, the radiation has been blueshifted. The blueshift is experienced by the
photons that propagate for a sufficiently long time inside the collapsing body.
Indeed, if we compute the curves of $g(\lambda)$ and $r(\lambda)$ for the photons
with $b=0$, we find the picture in Fig.~\ref{f3}.

\begin{figure*}
\begin{center}
\hspace{-0.5cm}
\includegraphics[type=pdf,ext=.pdf,read=.pdf,width=7.8cm]{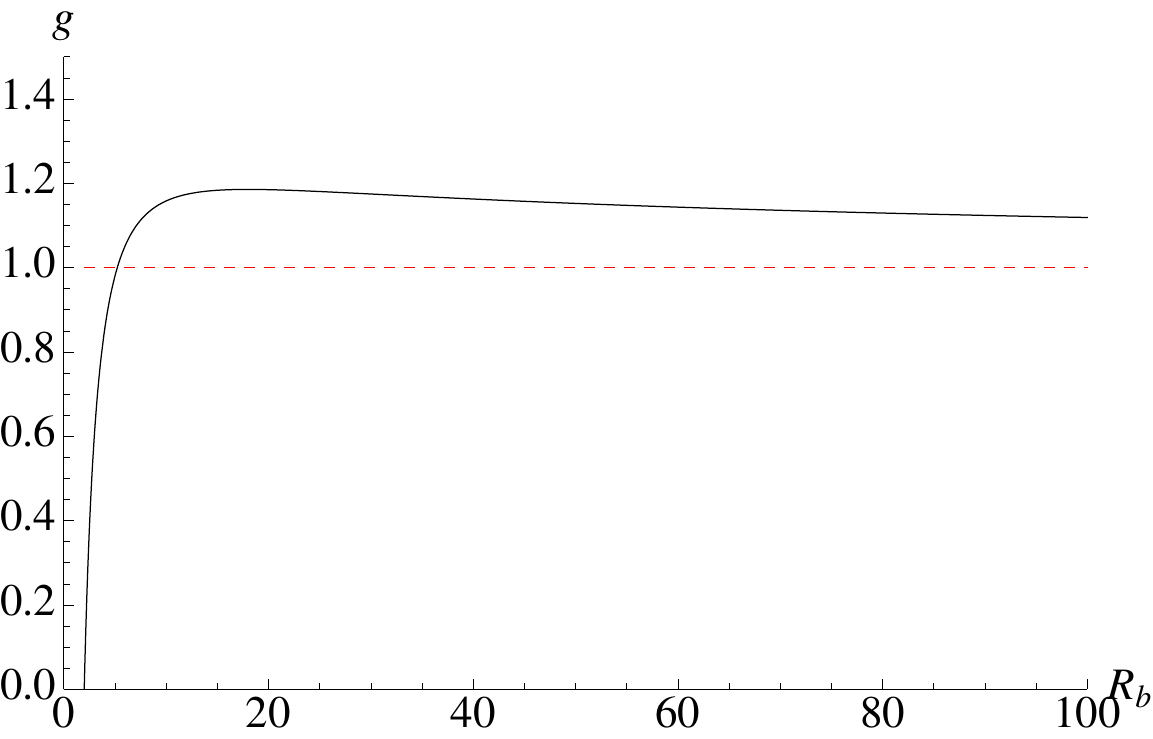}
\end{center}
\caption{Gravitational redshift $g$ as a function of $R_{\rm b}$, for a photon emitted 
on the surface of the collapsing body with pure radial negative velocity ($c = -1$). 
As the time of propagation inside the collapsing body can be long, the photon may 
be detected by the distant observer with a frequency $\nu_{\rm obs} > \nu_\star$, i.e. 
the photon is blueshifted. See the text for more details.}
\label{f4}
\end{figure*}

To check that the phenomenon is real and not the product of a numerical error,
it is convenient to do the analytical calculation in a simple example, namely the Oppenheimer-Snyder (OS) homogeneous case \cite{OS}. 
We consider in this context a photon emitted inside the collapsing body 
with pure radial negative velocity; that is, its impact parameter is $b=0$.
Let $t_1$ and $t_2$ be, respectively, the time of emission and of departure from
the interior. We can then exploit the fact that $ds^2_{\rm in} = 0$ and find a 
relation between the scale factor at the time $t_1$ and $t_2$
\be
r(t_2) - r(t_1) = \int_{t_1}^{t_2} \frac{d\tau}{a(\tau)} =\frac{2}{\sqrt{M_0}} 
\left[\sqrt{a(t_1)} - \sqrt{a(t_2)}\right] \, , \nonumber
\ee
where we have used Eq.~(\ref{a1}) and the fact that in the homogeneous case 
$M = M_0$. Since the photon exits the body at the boundary, $r(t_2) = r_{\rm b}$.
We can instead write the radius of the emission as $r(t_1)= c r_b$, with $c \in [-1,1]$.
For $c = 1$, $r(t_1)$ and $r(t_2)$ coincide and the photon is emitted from the
boundary to the exterior, so it only propagates in the Schwarzschild vacuum spacetime.
For $c = 0$, the photon is emitted from the center of the cloud. For $c=-1$, the photon 
is emitted from the boundary inwards, so it traverses the whole cloud before reaching
the boundary on the other side. If we call $R_i = a(t_i) r_{\rm b}$, we have
\be
r_{\rm b} - c r_{\rm b} = \frac{\sqrt{2} r_{\rm b}}{\sqrt{M_{\rm Sch}}} 
\left[\sqrt{R_1} - \sqrt{R_2}\right] \, ,
\ee
since $2 M_{\rm Sch} = r^3_{\rm b} M_0$. As $R_2 = R_{\rm b}$, 
we can write
\be
R_1 = \left[\sqrt{R_{\rm b}} + \sqrt{\frac{M_{\rm Sch}}{2}}(1 - c)\right]^2 \, .
\ee
In this simple case, the redshift factor $g$ is given by
\begin{widetext}
\be
g &=& \frac{E}{ \left(\frac{dt}{d\lambda}\right)_1} =
\frac{E}{ \frac{a(t_1)}{a(t_2)} \left(\frac{dt}{d\lambda}\right)_{r_{\rm b}}} =
\left[ \frac{R_1}{R_{\rm b}}\frac{1}{E}\left(\frac{dt}{d\lambda}\right)_{r_{\rm b}}\right]^{-1}=
\frac{\left[\sqrt{R_{\rm b}} + \sqrt{\frac{M_{\rm Sch}}{2}}(1 - c)\right]^2}{R_{\rm b}}
\frac{R_{\rm b} - 2 M_{\rm Sch}}{R_{\rm b}}
\frac{\sqrt{R_{\rm b}}}{\sqrt{R_{\rm b}} + \sqrt{2 M_{\rm Sch}}} \, .
\ee 
\end{widetext}
where the factor $R_1/R_{\rm b}>1$ is the blueshift produced inside the collapsing
object and the factor $(1/E)(dt/d\lambda)_{r_{\rm b}}$ is the redshift~(\ref{eq-tra}) experienced
during the propagation in the Schwarzschild spacetime, which includes the
Doppler redshift caused by the motion of the surface of the collapsing body. 
For $c = -1$, $g$ is not a monotonically decreasing function (as, for instance, it
is in the case $c=0$ or $c=1$). As shown in Fig.~\ref{f4},
$g$ goes to 0 for $R_{\rm b} \rightarrow 2 M_{\rm Sch}$
(photons are infinitely redshifted) and to 1 for $R_{\rm b} \rightarrow \infty$ (no shift), 
but for $R_{\rm b} > (3 + \sqrt{5}) \; M_{\rm Sch}$, $g > 1$ (photons are blueshifted), 
and the maximum is $g = 32/27 \approx 1.185$ at $R_{\rm b} = 18 \; M_{\rm Sch}$.

{\it Conclusions ---}
In this note, we have considered the gravitational collapse of a spherically symmetric
cloud of dust and we have computed the spectrum of the radiation emitted by this
body for a simple emissivity function. 
The problem is interesting if one wishes to test the observable nature of the ``naked singularities''
that arise in some theoretical models~\cite{kmb}
(see~\cite{review} and references therein for a review on gravitational collapse 
and~\cite{nakao,nakao2} for issues related to photons emitted from the singularities).
The exterior metric is described by the
Schwarzschild solution, while the geometry of the interior spacetime is described 
by the LTB metric. The photons are blueshifted when they
propagate inside the collapsing object, and redshifted when they propagate in
the vacuum, from the surface of the body to the detector of the distant observer. 
The counterintuitive phenomenon is that a non-negligible fraction of the radiation reaching 
far-away observers may be blueshifted. 
The key point is that some radiation, produced in the interior with frequency 
$\omega_i$, can propagate inside the cloud for a sufficiently long time to 
reach the boundary with blueshifted frequency $\omega_b>\omega_i$ so 
that the subsequent redshift occurring in the vacuum will still allow the photon 
to reach the distant observer with an energy larger than the initial one, and 
therefore with a frequency $\omega_f>\omega_i$ (though the usual redshift 
in vacuum still necessarily implies $\omega_f<\omega_b$).
Unfortunately, a direct observational confirmation of this effect seems to be 
very unlikely. A star is never transparent. For instance, even the core of a 
supernova is opaque to photons as well as to neutrinos. So this effect does 
not seem to have direct observational consequences.

We thank Ken-ichi Nakao for useful comments and suggestions.
This work was supported by the NSFC grant No.~11305038,
the Shanghai Municipal Education Commission grant for Innovative 
Programs No.~14ZZ001, the Thousand Young Talents Program, 
and Fudan University.


\end{document}